\documentclass[12pt,reqno]{amsart}
\usepackage{amssymb,latexsym,amsmath}
\usepackage{amscd,amsfonts}
\usepackage{epsfig}

\newtheorem{theorem}{Theorem}
\newtheorem{lemma}{Lemma}

\newtheorem{corollary}{Corollary}

\theoremstyle{definition}
\newtheorem{remark}{Remark}
\newtheorem{example}{Example}
\numberwithin{equation}{section}

\begin{document}
\title{Weinhold'length in an isentropic Ideal and quasi-Ideal Gas}
\author{Manuel Santoro}\address{Department of Mathematics and
Statistics, Portland State University, PO Box 751, Portland, OR
97207-0751, USA}
 \email{emanus@pdx.edu}
 Dedicated to: Gianluca, Paola and Marco Loiacono
\begin{abstract}
In this paper we study thermodynamic length of an isentropic Ideal
and quasi-Ideal Gas using Weinhold metric in a two-dimensional
state space. We give explicit relation between length at constant
entropy and work.
\end{abstract}\bigskip
 \maketitle
\section{Introduction}
\par
Weinhold introduced a metric$^{12}$ $\eta_{ij}$ in the space of
thermodynamic states as second derivative of internal energy with
respect to extensive variables $X_{i}$ and $X_{j}$, namely
$\eta_{ij}=\frac{\partial^{2}U}{\partial{X_{i}X_{j}}}$ with
$i,j=1,..,n$. In a general setting, constitutive relation
$U=U(X_{1},...,X_{n})$ represents the energy surface in which, for
example, $X_{1}=S$, $X_{2}=V$,etc., where S is the entropy and V
is volume of our system. Such a metric gives us a way to define
distances and angles and, therefore, it enables us to study the
geometry of the surface. Several important questions were
considered both from a geometrical and a physical point of view.
In this manuscript we are concerned with one of them: what is the
meaning of thermodynamic length?
\par
Weinhold's metric $\eta_{ij}$ was used by P.Salamon$^{5,6,7,8}$,
R.S.Berry$^{5,7,8}$, J.Nulton$^{6,8}$, E.Ihrig$^{6}$, and
others$^{7,9}$ to study such a question. The \textit{local}
meaning of $\eta_{ij}$ is the distance$^{7}$ between the energy
surface and the linear space tangent to this surface at some point
where $\eta_{ij}$ is evaluated. Indeed, let's denote by
coordinates $(X^{0}_{1},...,X^{0}_{n})$ a particular energy state.
The tangent space is attached to the energy surface at point
$(U_{0},X^{0}_{1},...,X^{0}_{n})$. If we move away a little to a
new energy state $(X_{1},...,X_{n})$ then the distance between the
point on the surface $(U,X_{1},...,X_{n})$ and the tangent space
is the \textit{availability}$^{5,7}$ or the available work$^{5}$
of the system. This interpretation is only local since it requires
just small displacements, like for fluctuations, from the given
point $(U_{0},X^{0}_{1},...,X^{0}_{n})$ on the surface.
\par
On the other hand, we could study thermodynamic length taking the
metric $\eta_{ij}$ \textit{globally}. In this situation we
consider a path $\phi$ on the energy surface between two states
$a_{0}$ and $a_{1}$ and study the length of the path
\par
\[
L_{a_{0}a_{1}}=\int^{a_{1}}_{a_{0}}[\sum_{i,j}{\eta_{ij}dX_{i}dX_{j}}]^{\frac{1}{2}}
\]
\par
It was shown$^{7}$ that L, in general, represents the change in
mean molecular velocity depending on the particular nature of the
thermodynamic process defining the path $\phi$ and that its
dimension is square root of energy. But thermodynamic length was
explicitly studied$^{7}$ just in the Ideal case. In particular, it
was found that, for a reversible adiabatic Ideal Gas from state
$(p_{0},V_{0})$ to state $(p_{1},V_{1})$, length represents the
change$^{7}$ in flow velocity of a gas undergoing an isentropic
expansion, like in rarefaction waves, and it is given by
\par
\[
L^{(s)}=\frac{2}{\gamma-1}\sqrt{\gamma{p_{0}}V_{0}}[1-(\frac{p_{1}}{p_{0}})^{\frac{\gamma-1}{2\gamma}}]
\]
\par
with $\gamma=\frac{C_{p}}{C_{v}}$.
\par
Our previous work$^{9}$ and the present paper are further studies
on global thermodynamic equilibrium and length.  We explicitly
found$^{9}$ a relation between thermodynamic length and work for
an isentropic Ideal and quasi-Ideal Gas along isotherms, namely
\par
\[
L^{(s)}=\sqrt{\frac{1}{RT}}W
\]
\par
Naturally, we thought of thermodynamic length as a measure of the
amount of work done by the system along isotherms. But its
interpretation in relation with work turned out to be richer than
what we had recognized. Indeed, we realized that such a case was
the trivial one and, therefore, we give, in this manuscript, a
generalization of that relation no longer at constant temperature.
In particular, we found that thermodynamic length of an isentropic
Ideal or quasi-Ideal Gas measures the difference of the square
roots of the energies of  two given states. Naturally, if there is
no work received or done by such a system then the length of the
path $\phi$ is zero.
\par
\section{Thermodynamic length with Weinhold metric}
\par
Consider the thermodynamic length between two states $a_{0}$ and
$a_{1}$ of our system
\par
\begin{equation}
L_{a_{0}a_{1}}=\int^{a_{1}}_{a_{0}}[\sum_{i,j}{\eta_{ij}dX_{i}dX_{j}}]^{\frac{1}{2}}
\end{equation}
\par
where $\eta_{ij}$ are elements of the thermodynamic metric and
$X_{i}$ represent independent coordinates in thermodynamic state
space. We shall see that thermodynamic length of an isentropic
Ideal Gas with two degrees of freedom is related to the concept of
work of a reversible process.
\par
\mathstrut \par Let's consider constitutive relation $u=u(s,v)$
where u is the molar internal energy, s is the molar entropy and v
is the molar volume. s and v are the two independent variables.
Weinhold metric is given by$^{3}$
\par
\begin{equation}
\eta_{ij}=\frac{1}{c_{v}}
\begin{pmatrix}
 T & -\frac{T\alpha}{k_{T}}\\
 -\frac{T\alpha}{k_{T}} & \frac{c_{p}}{vk_{T}}
\end{pmatrix}
\end{equation}
\par
where
\par
\begin{enumerate}
\item $c_{v}$ is the molar heat capacity at constant volume:
\[
c_{v}=T(\frac{\partial s}{\partial T})_{v}\qquad,
\]
\item $c_{p}$ is the molar heat capacity at constant pressure:
\[
c_{p}=T(\frac{\partial s}{\partial T})_{p}\qquad,
\]
\item $\alpha$ is the thermal coefficient of expansion:
\[
\alpha=\frac{1}{v}(\frac{\partial v}{\partial T})_{p}\qquad,
\]
\item $\kappa_{T}$ is the isothermal compressibility:
\[
\kappa_{T}=-\frac{1}{v}(\frac{\partial v}{\partial p})_{T}\qquad.
\]
\end{enumerate}
\par
Thermodynamic length with Weinhold's metric is given by$^{7}$
\par
\begin{equation}
L=\int{[\frac{T}{c_{v}}(ds)^{2}-2\frac{T\alpha}{c_{v}\kappa_{T}}dsdv+\frac{c_{p}}{vc_{v}\kappa_{T}}(dv)^{2}]^{\frac{1}{2}}}
\end{equation}
\par
and, if molar entropy and molar volume are given parametrically as
$s=s(\xi)$, $v=v(\xi)$, then we have$^{7}$
\par
\begin{equation}
L=\int^{\xi_{f}}_{\xi_{i}}[\frac{T}{c_{v}}(\frac{ds}{d\xi})^{2}-2\frac{T\alpha}{c_{v}\kappa_{T}}\frac{ds}{d\xi}\frac{dv}{d\xi}+\frac{c_{p}}{vc_{v}\kappa_{T}}(\frac{dv}{d\xi})^{2}]^\frac{1}{2}d\xi
\end{equation}
\par
Consider thermodynamic length at constant entropy given by$^{9}$
\par
\[
L^{s}=\int{\sqrt{\frac{c_{p}}{c_{v}vk_{T}}}dv}=\int{\sqrt{\eta_{22}}dv}
\]
\par
In the Ideal Gas case, it becomes
\par
\begin{equation}
L^{s}=\int{\sqrt{\frac{c_{p}p}{c_{v}v}}dv}
\end{equation}
\par
since $k_{T}=\frac{1}{p}$.
\par
\section{Thermodynamic length in an isentropic TD system with constant heat capacity}
\par
Let's consider constant molar heat capacity at constant volume
$c_{v}=T(\frac{\partial s}{\partial T})_{v}$. It easily follows
that $(\frac{\partial T}{\partial s})_{v}=\frac{T}{c_{v}}$ and,
since $T=(\frac{\partial u}{\partial s})_{v}$, we get the
following equation$^{11}$
\par
\begin{equation}
\frac{\partial ^{2}u}{\partial
s^{2}}-\frac{1}{c_{v}}\frac{\partial u}{\partial s}=0.
\end{equation}
\par
Integrating twice we get first that
\[
\frac{\partial u}{\partial s}=\frac{u}{c_{v}}+f_{2}(v),
\]
with an arbitrary function $f_{2}(v)$ and, second, the fundamental
constitutive law in the form$^{11}$
\begin{equation}
u(s,v)=f_{1}(v)e^{\frac{s-s_{0}}{c_{v}}}-c_{v}f_{2}(v),
\end{equation}
with another arbitrary function $f_{1}(v)$.
\par
\begin{example}
\textbf{Ideal Gas}
\par
It is known$^{1}$ that for an Ideal Gas
\[
s=s_{0}+c_{v}\ln{(\frac{u}{u_{0}})}+R\ln{(\frac{v}{v_{0}})}
\]
\par
Let $u_{0}=v_{0}=1$ for simplicity. Then, solving for the internal
energy, we get
\par
\begin{equation}
u=v^{-\frac{R}{c_{v}}}e^{\frac{s-s_{0}}{c_{v}}}
\end{equation}
\par
Therefore, considering $(3.2)$, we get the Ideal Gas case if we
set $f_{1}(v)=v^{-\frac{R}{c_{v}}}$ and $f_{2}(v)=0$.$^{11}$
\end{example}
\par
\begin{example}
\textbf{Quasi-Ideal Gas}
\par
We also consider the quasi-Ideal case$^{9}$ in which
$f_{1}(v)=(v-b)^{-\frac{R}{c_{v}}}$ and $f_{2}(v)=0$, with b
positive constant.
\end{example}
\par
\begin{example}
\textbf{Van der Waals Gas}
\par
The entropy function of the Van der Waals Gas is given by$^{1}$
\par
\[
s=s_{0}+R\ln{[(v-b)(u+\frac{a}{v})^{\frac{c_{v}}{R}}]}
\]
\par
where $a$ and $b$ are positive constants.
\par
Then, solving for u, we get
\par
\begin{equation}
u=(v-b)^{-\frac{R}{c_{v}}}e^{\frac{s-s_{0}}{c_{v}}}-\frac{a}{v}
\end{equation}
\par
Therefore, considering the general case with heat capacity
constant $(3.2)$, we get the Van der Waals Gas case if we set
$f_{1}(v)=(v-b)^{-\frac{R}{c_{v}}}$ and
$f_{2}(v)=\frac{a}{c_{v}v}$.$^{11}$
\end{example}
\par
Considering equation $(3.2)$, we have that
$(\frac{\partial^{2}u}{\partial{v^{2}}})_{s}=f^{''}_{1}(v)e^{\frac{s-s_{0}}{c_{v}}}-c_{v}f^{''}_{2}(v)$,
where $f^{''}$ indicates second derivative with respect to molar
volume. Therefore, thermodynamic length at constant entropy can be
written as
\par
\begin{equation}
L^{s}=\int{\sqrt{\eta_{22}}dv}=\int{\sqrt{(\frac{\partial^{2}u}{\partial{v^{2}}})_{s}}dv}=\int{\sqrt{f^{''}_{1}(v)e^{\frac{s-s_{0}}{c_{v}}}-c_{v}f^{''}_{2}(v)}dv}
\end{equation}
\par
Now, since for an isentropic system
$dW=du=-pdv=[f^{'}_{1}(v)e^{\frac{s-s_{0}}{c_{v}}}-c_{v}f^{'}_{2}(v)]dv$,
where $dW$ is the infinitesimal work per unit mole and $f^{'}$ are
first derivatives with respect to v, then we have the following
result
\par
\begin{lemma}
\begin{equation}
(\frac{dL^{s}}{dv})^{2}=\frac{d^{2}W}{dv^{2}}
\end{equation}
\end{lemma}
\par
\subsection{Relation between "isentropic" length and work for an Ideal Gas}
\par
Here we'll derive a relation between the work $W$ as difference in
molar internal energy and thermodynamic length of a reversible
isentropic Ideal Gas.\par Since the molar entropy is constant,
consider $u_{2}=u(s,v_{2})$ and $u_{1}=u(s,v_{1})$. Since
$W=\Delta{u}=u_{2}-u_{1}$, then, considering $(3.2)$, we have, in
general, that
\par
\begin{equation}
W=u_{2}-u_{1}=[f_{1}(v_{2})-f_{1}(v_{1})]e^{\frac{s-s_{0}}{c_{v}}}-c_{v}[f_{2}(v_{2})-f_{2}(v_{1})]
\end{equation}
\par
Let's study, now, the Ideal case. Denote positive work as work
done on the system,
\par
\[
W_{in}=W>0
\]
\par
and negative work as work done by the system,
\par
\[
W_{out}=W<0
\]
\par
As mentioned in example $1$, $f_{1}(v)=v^{-\frac{R}{c_{v}}}$ and
$f_{2}(v)=0$. Therefore, letting $v_{1}<v_{2}$, we have the
following expressions,
\par
\begin{equation}
W_{in}=u_{1}-u_{2}=[v_{1}^{-\frac{R}{c_{v}}}-v_{2}^{-\frac{R}{c_{v}}}]e^{\frac{s-s_{0}}{c_{v}}}
\end{equation}
\par
\begin{equation}
W_{out}=u_{2}-u_{1}=[v_{2}^{-\frac{R}{c_{v}}}-v_{1}^{-\frac{R}{c_{v}}}]e^{\frac{s-s_{0}}{c_{v}}}
\end{equation}
\par
\begin{remark}
Note that
$p=-(\frac{\partial{u}}{\partial{v}})_{s}=\frac{R}{c_{v}}v^{-\frac{c_{p}}{c_{v}}}e^{\frac{s-s_{0}}{c_{v}}}$.
Therefore, the work done by the system is given by
\par
\[
W_{out}=-\int^{v_{2}}_{v_{1}}{pdv}=-\frac{R}{c_{v}}e^{\frac{s-s_{0}}{c_{v}}}\int^{v_{2}}_{v_{1}}{v^{-\frac{c_{p}}{c_{v}}}dv}=[v_{2}^{-\frac{R}{c_{v}}}-v_{1}^{-\frac{R}{c_{v}}}]e^{\frac{s-s_{0}}{c_{v}}}
\]
\end{remark}
\par
\mathstrut
\par
On the other hand, requiring length to be always positive or equal
to zero and considering $(3.5)$, we have
\par
\begin{equation}
L^{s}=\int^{v_{2}}_{v_{1}}{\sqrt{\frac{Rc_{p}}{c^{2}_{v}}v^{-\frac{R}{c_{v}}-2}e^{\frac{s-s_{0}}{c_{v}}}}dv}=2\sqrt{\frac{c_{p}}{R}}[v_{1}^{-\frac{R}{2c_{v}}}-v_{2}^{-\frac{R}{2c_{v}}}]e^{\frac{s-s_{0}}{2c_{v}}}>0
\end{equation}
\par
\begin{equation}
-L^{v}=\int^{v_{1}}_{v_{2}}{\sqrt{\frac{Rc_{p}}{c^{2}_{v}}v^{-\frac{R}{c_{v}}-2}e^{\frac{s-s_{0}}{c_{v}}}}dv}=2\sqrt{\frac{c_{p}}{R}}[v_{2}^{-\frac{R}{2c_{v}}}-v_{1}^{-\frac{R}{2c_{v}}}]e^{\frac{s-s_{0}}{2c_{v}}}<0
\end{equation}
\par
From these expressions of work and length follows,
\par
\begin{theorem}
\par
\begin{equation}
W_{in}=\frac{RL^{s}}{4c_{p}}[L^{s}+4\sqrt{\frac{c_{p}}{R}u_{2}}]
\end{equation}
where $u_{2}=v^{-\frac{R}{c_{v}}}_{2}e^{\frac{s-s_{0}}{c_{v}}}$ is
the internal energy of an isentropic Ideal gas evaluated at molar
volume $v_{2}$, or, equivalently,
\par
\begin{equation}
(L^{s})^{2}+4\sqrt{\frac{c_{p}}{R}u_{2}}L^{s}-4\frac{c_{p}}{R}W_{in}=0
\end{equation}
\end{theorem}
\par
It easily follows that
\par
\begin{equation}
W_{out}=-W_{in}=-\frac{RL^{s}}{4c_{p}}[L^{s}+4\sqrt{\frac{c_{p}}{R}u_{2}}]
\end{equation}
or, equivalently,
\par
\begin{equation}
(L^{s})^{2}+4\sqrt{\frac{c_{p}}{R}u_{2}}L^{s}+4\frac{c_{p}}{R}W_{out}=0
\end{equation}
\par
We also have the following two corollaries,
\par
\begin{corollary}
\begin{equation}
L^{s}=2\sqrt{\frac{c_{p}}{R}}[\sqrt{u_{2}+W_{in}}-\sqrt{u_{2}}]=2\sqrt{\frac{c_{p}}{R}}[\sqrt{u_{1}}-\sqrt{u_{2}}]
\end{equation}
\end{corollary}
\par
which is equivalent to
\par
\begin{equation}
-L^{s}=2\sqrt{\frac{c_{p}}{R}}[\sqrt{u_{2}}-\sqrt{u_{2}-W_{out}}]=2\sqrt{\frac{c_{p}}{R}}[\sqrt{u_{2}}-\sqrt{u_{1}}]
\end{equation}
\par
\begin{corollary}
\par
\begin{equation}
 \pm{L^{s}}=0\qquad iff\qquad W=0
\end{equation}
\end{corollary}
\par
\begin{remark}
Note that these results can be applied to the quasi-Ideal case by
substituting $v_{i}-b$ to $v_{i}$ for $i=1,2$.
\end{remark}
\par
\begin{remark}
Consider equation $(3.2)$. Then, temperature T is given by
$T(s,v)=(\frac{\partial u}{\partial
s})_{v}=\frac{f_{1}(v)}{c_{v}}e^{\frac{s-s_{0}}{c_{v}}}$ for a
generic two-dimensional thermodynamic system at constant heat
capacity. It is evident that, in an isentropic thermodynamic
system, temperature is constant if and only if $f_{1}(v)$ is
constant. Therefore, considering equation $(3.7)$, we have that
work $W$ is zero for an Ideal and quasi-Ideal Gas along isotherms.
Thus, length is zero. In particular, we make clear that for an
Ideal and quasi-Ideal Gas, thermodynamic length in an isentropic
system along isotherms$^{9}$ is zero,i.e.
\par
\begin{equation}
L^{s}=\sqrt{\frac{1}{RT}}W=0
\end{equation}
\end{remark}
\par
\section{Conclusions}
\par
We gave a physical interpretation of thermodynamic length in a
simple isentropic Ideal and quasi-Ideal Gas with two degrees of
freedom .
\par

\end{document}